\documentclass[preprint,showpacs,preprintnumbers,amsmath,amssymb,superscriptaddress,12pt,graphicx]{revtex4}

\usepackage{graphicx}
\usepackage{dcolumn}
\usepackage{bm}
\usepackage{amssymb}

\renewcommand{\H}{{\cal H}}
\renewcommand{\L}{{\cal L}}

\newcommand{\be}{\begin{equation}}
\newcommand{\ee}{\end{equation}}
\newcommand{\bee}{\begin{eqnarray}}
\newcommand{\een}{\end{eqnarray}}
\newcommand{\ba}{\begin{eqnarray}}
\newcommand{\ea}{\end{eqnarray}}

\begin{document}

\title{The State Equation of Yang-Mills Field Dark Energy Models}

\author{Wen Zhao}
\email{wzhao7@mail.ustc.edu.cn}\affiliation{ Astrophysics Center, University of Science and Technology of China, Hefei, Anhui, China}
\author{Yang Zhang}
\affiliation{ Astrophysics Center, University of Science and Technology of China, Hefei, Anhui, China}

%\date{\today}

%%%%%%%%%%%%%%%%%%%%%%%%%%%%%%%%%%%%%%%%%%%%%%%%%%%%%%%%%%%%%%%%%%%%%%%%%%%%%%%%%%%
%%%%%%%%%%%%%%%%%%%%%%%%%%%%%%%%%%  ABSTRACT  %%%%%%%%%%%%%%%%%%%%%%%%%%%%%%%%%%%%%%%%%%
%%%%%%%%%%%%%%%%%%%%%%%%%%%%%%%%%%%%%%%%%%%%%%%%%%%%%%%%%%%%%%%%%%%%%%%%%%%%%%%%%%%

\begin{abstract}
In this paper, we study the possibility of building Yang-Mills(YM)
field dark energy models with equation of state (EoS) crossing -1,
and find that it can not be realized by the single YM field
models, no matter what kind of lagrangian or initial condition.
But the states of $-1<\omega<0$ and $\omega<-1$ all can be
naturally got in this kind of models. The former is like a
quintessence field, and the latter is like a phantom field. This
makes that one can build a model with two YM fields, in which one
with the initial state of $-1<\omega<0$, and the other with
$\omega<-1$. We give an example model of this kind, and find that
its EoS  is larger than -1 in the past and less than -1 at the
present time. We also find that this change must be from
$\omega>-1$ to $<-1$, and it will go to the critical state of
$\omega=-1$ with the expansion of the Universe, which character is
same with the single YM field models, and the Big Rip is naturally
avoided.
\end{abstract}

%%%%%%%%%%%%%%%%%%%%%%%%%%%%%%%%%%%%%%%%%%%%%%%%%%%%%%%%%%%%%%%%%%%%%%%%%%%%%%%%%%%
%%%%%%%%%%%%%%%%%%%%%%%%%%%%%%%%%%%%%%%%%%%%%%%%%%%%%%%%%%%%%%%%%%%%%%%%%%%%%%%%%%%


\maketitle

%%%%%%%%%%%%%%%%%%%%%%%%%%%%%%%%%%%%%%%%%%%%%%%%%%%%%%%%%%%%%%%%%%%%%%%%%%%%%%%%%%%
%%%%%%%%%%%%%%%%%%%%%%%%%%%%%%%%%%  SECTION 1   %%%%%%%%%%%%%%%%%%%%%%%%%%%%%%%%%%%%%%%%%%
%%%%%%%%%%%%%%%%%%%%%%%%%%%%%%%%%%%%%%%%%%%%%%%%%%%%%%%%%%%%%%%%%%%%%%%%%%%%%%%%%%%

\section{Introduction}

Recent observations on the Type Ia Supernova (SNIa)\cite{sn},
Cosmic Microwave Background Radiation (CMB)\cite{map} and Large
Scale Structure (LSS)\cite{sdss} all suggest that the Universe
mainly consists of dark energy (73\%), dark matter (23\%) and
baryon matter (4\%). How to understand the physics of the dark
energy is an important mission in the modern cosmology, which has
the EoS of $\omega<-1/3$, and leads to the recent accelerating
expansion of the Universe. Several scenarios have been put forward
as a possible explanation of it. A positive cosmological constant
is the simplest candidate, but it needs the extreme fine tuning to
account for the observed accelerating expansion of the Universe.
This fact has led to models where the dark energy component varies
with time, such as quintessence models\cite{quint}, which assume
the dark energy is made of a single (light) scalar field.  Despite
some pleasing features, these models are not entirely
satisfactory, since in order to achieve $\Omega_{de}\sim\Omega_m$
(where $\Omega_{de}$ and $\Omega_m$ are the dark energy and matter
energy densities at present, respectively) some fine tuning is
also required. Many other possibilities have been considered for
the origin of this dark energy component such as a scalar field
with a non-standard kinetic term and k-essence models\cite{k}, it
is also possible to construct models which have the EoS of
$\omega=p/\rho<-1$, the so-called phantom\cite{phantom}. Some
other models such as the generalized Chaplygin gas (GCG)
models\cite{GCG}, the vector field models\cite{vec} also have been
studied by a lot of authors. Although these models achieve some
success, some problems also exist. One essential to understand the
nature of the dark energy is to detect the value and evolution of
its EoS.  The observation data shows that the cosmological
constant is a good candidate\cite{seljak}, which has the effective
equation $p=-\rho$, $i.e.$ $\omega\equiv-1$. However, there is an
evidence to show that the dark energy might evolve from
$\omega>-1$ in the past to $\omega<-1$ today, and cross the
critical state of $\omega=-1$ in the intermediate
redshift\cite{trans}. If such a result holds on with accumulation
of observational data, this would be a great challenge to the
current models of dark energy. It is obvious that the cosmological
constant as a candidate will be excluded, and dark energy must be
dynamical. But the normal models such as the quintessence fields,
only can give the state of $-1<\omega<0$. Although the k-essence
models and the phantom models can get the state of $\omega<-1$,
but the behavior of $\omega$ crossing $-1$ can not be realized,
and all these will lead to theoretical problem in field theory. To
answer this crossing phenomenon of $\omega$, a lot of people have
advised some more complex models, such as the quintom
models\cite{quintom,quintom1}, which is made of a quintessence
field and a phantom field. The model with higher derivative term
has been suggested in Ref.\cite{lmz}, which also can get from
$\omega>-1$ to $\omega<-1$, but it also will lead to theoretical
difficulty in field theory.

We have advised that the YM field\cite{Zhang,zhao} can be used to
describe the dark energy. There are two major reason that prompt
us to study this system. First the normal scalar models the
connection of field to particle physics models has not been clear
so far. The second reason is that the weak energy condition can
not be violated by the field. The YM field we have advised has the
desired interesting featured: the YM field are the indispenable
cornerstone to any particle physics model with interactions
mediated by gauge bosons, so it can be incorporated into a
sensible unified theory of particle physics. Besides, the EoS of
matter for the effective YM condensate is different from that of
ordinary matter as well as the scalar fields, and the state of
$-1<\omega<0$ and $\omega<-1$ can also be naturally realized. But
if it is possible to build a YM field model with EoS crossing
$-1$? In this paper, we focus on this topic. First we consider the
YM field with a general lagrangian, and find the state of
$\omega\sim-1$ is easily realized, as long as it satisfies some
constraint. From the kinetic equation of the YM field, we find
that $\omega+1\propto a^{-2}$ with the expansion of the Universe.
But no matter what kind of lagrangian and initial condition we
choose, this model can not get a behavior of $\omega$ crossing
$-1$. But it can be easily got in the models with two YM fields,
one with the initial condition of $\omega>-1$, which is like a
quintessence field, and the other with $\omega<-1$ like a phantom
field.

This paper is organized as follows. In section 2 we discuss the
general YM field model, and study the evolution of its EoS by
solving its kinetic equation. But we find that this kind of model
can not get the state of $\omega$ crossing $-1$. Then we study the
two YM fields model in section 3, and solve the evolution of
$\omega$ with scale factor for an example model. We find that
$\omega$ crossing $-1$ can be easily realized in this model, which
is very like the quintom models. At last, we have a conclusion and
discussion in section 4.

%%%%%%%%%%%%%%%%%%%%%%%%%%%%%%%%%%%%%%%%%%%%%%%%%%%%%

\section{ Single YM Field Model}

In the Ref.\cite{zhao}, we have discussed the EoS of the YM field
dark energy models, which has the effective
lagrangian\cite{pagels, adler}
 \be
 \L_{eff}=F/2g^2.
 \ee
here $F=-(1/2)F^a_{\mu\nu}F^{a\mu\nu}$ plays the role of the order
parameter of the YM condensate, and $g$ is the running coupling
constant which, up to 1-loop order, is given by
 \be
 \frac{1}{g^2}=b\ln|\frac{F}{\kappa^2}-1|.
 \ee
Thus the effective lagrangian is
 \be
 \L_{eff}=\frac{b}{2}F\ln|\frac{F}{e\kappa^2}|, \label{L}
 \ee
where $e\simeq2.72$. $b=11N/24\pi^2$ for the generic gauge group
$SU(N)$ is the Callan-Symanzik coefficient\cite{Pol}, $\kappa$ is
the renormalization scale with the dimension of squared mass, the
only model parameter. The attractive features of this effective YM
action model include the gauge invariance, the Lorentz invariance,
the correct trace anomaly, and the asymptotic
freedom\cite{pagels}. With the logarithmic dependence on the field
strength, $\L_{eff}$ has a form similar to he Coleman-Weinberg
scalar effective potential\cite{coleman}, and the Parker-Raval
effective gravity lagrangian\cite{parker}.

It is straightforward to extend this model to the expanding
Robertson-Walker (R-W) spacetime. For simplicity we will work in a
spatially flat R-W spacetime with a metric
 \be
 ds^2=a^2(\tau)(d\tau^2-\gamma_{ij}dx^idx^j),\label{me}
 \ee
where we have set the speed of light $c\equiv1$,
$\gamma_{ij}=\delta^i_j$ denoting the background space is flat,
and $\tau=\int(a_0/a)dt$ is the conformal time. Consider the
dominant YM condensate minimally coupled to the general relativity
with the effective action,
 \be
 S=\int \sqrt{-\tilde{g}}~[-\frac{R}{16\pi G}+\L_{eff}] ~d^{4}x,
 \label{S}
 \ee
where $\tilde{g}$ is the determinant of the metric $g_{\mu\nu}$.
By variation of $S$ with respect to the metric $g^{\mu\nu}$, one
obtains the Einstein equation $G_{\mu\nu}=8\pi GT_{\mu\nu}$, where
the energy-momentum tensor is given by
 \be
 T_{\mu\nu}=\sum_{a}~\frac{g_{\mu\nu}}{4g^2}F_{\sigma\delta}^a
 F^{a\sigma\delta}+\epsilon F_{\mu\sigma}^aF^{a\sigma}_{\nu}.
 \label{T}
 \ee
The dielectric constant is defined by
$\epsilon=2\partial\L_{eff}/\partial F$, and in this one-loop
order it is given by
 \be
 \epsilon=b\ln|\frac{F}{\kappa^2}|.\label{epsilon}
 \ee
This energy-momentum tensor is the sum of the several different
energy-momentum tensors of the vectors,
$T_{\mu\nu}=\sum_a~^{(a)}T_{\mu\nu}$, neither of which is of
prefect-fluid form, which can make the YM field being anisotropy.
This is one of the most important character of the vector field
dark energy models\cite{vec}.  If it is true and this anisotropy
YM field is dominant in the Universe, this will make the Universe
being anisotropy, one would expect an anisotropy expansion of the
Universe, in conflict with the significant isotropy of the
CMB\cite{isotropy}. But on the other hand there also appear to be
hints of statistical anisotropy in the CMB
perturbations\cite{fluctuate}. But here we only consider the other
case. For keeping the total energy-momentum tensor $T_{\mu\nu}$ is
homogeneous and isotropic, here we assume the gauge fields are the
functions of only time $t$, and
$A_{\mu}=\frac{i}{2}\sigma_aA_{\mu}^a(t)$ (here $\sigma_a$ are the
Pauli's matrices) are given by $A_0=0$ and $A_i^a=\delta_i^aA(t)$.
Define the YM field tensor as usual:
 \be
 F^{a}_{\mu\nu}=\partial_{\mu}A_{\nu}^a-\partial_{\nu}A_{\mu}^a+f^{abc}A_{\mu}^{b}A_{\nu}^{c},
 \ee
where $f^{abc}$ is the structure constant of gauge group and
$f^{abc}=\epsilon^{abc}$ for the $SU(2)$ case. This tensor can be
written in the form with the electric and magnetic field as
 \be
 F^{a\mu}_{~~\nu}=\left(
 \begin{array}{cccc}
      0 & E_1 & E_2 & E_3\\
     -E_1 & 0 & B_3 & -B_2\\
     -E_2 & -B_3 & 0 & B_1\\
     -E_3 & B_2 & -B_1 & 0
 \end{array}
 \right).
 \ee
It can be easily found that $E_1^2=E_2^2=E_3^2$, and
$B_1^2=B_2^2=B_3^2$. Thus $F$ has a simple form with $F=E^2-B^2$,
where $E^2=\sum_{i=1}^3E_i^2$ and $B^2=\sum_{i=1}^3B_i^2$. In this
case, each component of the energy-momentum tensor is
 \be
 ^{(a)}T_{\mu}^{0}=\frac{1}{6g^2}(B^2-E^2)\delta^{0}_{\mu}+\frac{\epsilon}{3}
 E^2\delta^{0}_{\mu},
 \ee
 \be
 ^{(a)}T_{j}^{i}=\frac{1}{6g^2}(B^2-E^2)\delta^i_j+\frac{\epsilon}{3}E^2\delta^i_j\delta^a_j
 -\frac{\epsilon}{3}B^2\delta^i_j(1-\delta^a_j).
 \ee
Although this tensor is not isotropic, its value along the $j=a$
direction is different from the one along the directions
perpendicular to it. Nevertheless, the total energy-momentum
tensor $T_{\mu\nu}=\sum_{a=1}^3~^{(a)}T_{\mu\nu}$ has isotropic
stresses, and the corresponding energy density and pressure are
given by (here we only consider the condition of
$B^2\equiv0$)\cite{zhao}
 \be
 \rho=\frac{E^2}{2}(\epsilon+b),~~~~p=\frac{E^2}{2}(\frac{\epsilon}{3}-b),
 \ee
and its EoS is
 \be
 \omega=\frac{\epsilon-3b}{3\epsilon+3b}.\label{13}
 \ee
It is easily found that at the critical point of $\epsilon=0$,
which follows that $\omega=-1$, the Universe is exact a de Sitter
expansion. Near this point, if $\epsilon<0$, we have $\omega<-1$,
and $\epsilon>0$ follows $\omega>-1$. So in these models, the EoS
of $0<\omega<-1$ and $\omega<-1$ all can be naturally realized.

For studying the evolution of this EoS, we should solve the YM
field equations, which is equivalent with solving the Einstein
equation\cite{zhao}. By variation of $S$ with respect to
$A_{\mu}^a$, one obtains the effective YM equations
 \be
 \partial_{\mu}(a^4\epsilon~
 F^{a\mu\nu})+f^{abc}A_{\mu}^{b}(a^4\epsilon~F^{c\mu\nu})=0.
 \label{F1}
 \ee
For we have assumed the YM condensate is homogeneous and
isotropic, from the definition of $F^{a}_{\mu\nu}$, it is easily
found that the $\nu=0$ component of YM equations is an identity
and the $i=1,2,3$ spatial components are:
 \be
 \partial_{\tau}(a^2\epsilon E)=0.
 \ee
If $\epsilon=0$, this equation is also an identity. When
$\epsilon\neq 0$, this equation follows that\cite{zhao},
 \be
 \beta~ e^{\beta/2}\propto a^{-2},\label{16}
 \ee
where we have defined $\beta\equiv\epsilon/b$, and used the
expression of $\epsilon$ in Eq.(\ref{epsilon}). In this equation,
the proportion factor can be fixed by the initial condition. This
is the main equation, which determines the evolution of $\beta$,
and $\beta$ directly relate to the EoS of the YM field. Combining
the Eqs.(\ref{13}) and (\ref{16}), one can obtains the evolution
of EoS in the YM field dark energy Universe. In Fig.[1], we plot
the the evolution of $\omega$ in the YM field dark energy models
with the present value $\omega_0=-1.2$ and $\omega_0=-0.8$, and
find that the former one is very like the evolution of the phantom
field, and the latter is like a quintessence field. They all have
same attractor solution with $\omega=-1$. So in these models, the
Big Rip is naturally avoided. This is the most attractive feature
of the YM field models.

In the Eq.(\ref{16}), the undetermined factor can be fixed by the
present value of EoS $\omega_0$, which must be determined by
observations on SNIa, CMB or LSS. In this paper, we will only show
that the observation of CMB power spectrum is an effective way to
determine it. The dark energy can influence the CMB temperature
anisotropy power spectrum (especially at the large scale) by the
integral Sachs-Wolfe(ISW) effect\cite{isw}. Consider the flat R-W
metric with the scalar perturbation in the conformal Newtonian
gauge,
 \be
 ds^2=a^{2}(\tau)[(1+2\phi)d\tau^2-(1-2\psi)\gamma_{ij}dx^idx^j].\label{metric}
 \ee
The gauge-invariant metric perturbation $\psi$ is the Newtonian
potential and $\phi$ is the perturbation to the intrinsic spatial
curvature. Always the background matters in the Universe are
perfect fluids without anisotropic stress, which follows that
$\phi=\psi$. So there is only one perturbation function $\phi$ in
the metric of (\ref{metric}), and its evolution is determined
by\cite{evolution}
 \be
 \phi''+3\H(1+\frac{p'}{\rho'})\phi'-\frac{p'}{\rho'}\nabla^2\phi
 +[(1+3\frac{p'}{\rho'})\H^2+2\H']\phi=4\pi
 Ga^2(\delta p-\frac{p'}{\rho'}\delta\rho),\label{Phi}
 \ee
where $\H\equiv a'/a$, and the $'$prime$'$ denotes $d/d\tau$. The
pressure $p=\sum_ip_i$, and energy density $\rho=\sum_i\rho_i$,
which should include the contribution of baryon, photon, neutron,
cold dark matter, and the dark energy. Especially at late time of
the Universe, the effect of the dark energy is very important. We
remind that the ISW effect stems from the time variation of the
metric perturbations,
 \be
 C_l^{ISW}\propto\int\frac{dk}{k}[\int_0^{\chi_{LSS}}d\chi~(\phi'+\psi')j_l(k\chi)]^2,
 \ee
where $\chi_{LSS}$ is the conformal distance to the last
scattering surface and $j_l$ the $l'$th spherical Bessel function.
The ISW effect occurs because photons can gain energy as they
travel through time-varying gravitational wells. One always solves
the CMB power spectrum in the numerical
methods\cite{cmbfast,camp}. In Fig.[2], we plot the CMB power
spectrum at large scale with these two kind of YM dark energy
models, where we have chosen the cosmological parameters as: the
Hubble parameter $h=0.72$, the energy density of baryon
$\Omega_bh^2=0.024$, and dark matter $\Omega_{dm}h^2=0.14$, the
reionization optical depth $\tau=0.17$, the spectrum index and
amplitude of the primordial perturbation spectrum being $n_s=0.99$
without running and $A=0.9$. Where we haven't consider the
perturbation of the dark energy. From this figure, one can find
that the values of the CMB power spectrums are very sensitively
dependent on $\omega_0$.
%are very different for YM dark energy models with different $\omega_0$.
Comparing with the $\Lambda$CDM model (which is equivalent with
the YM model with $\omega_0=-1$), the model with
$\omega_0=-0.8>-1.0$, which is like the quintessence field model,
the CMB spectrums have smaller values, especially at scale of
$l<10$, the difference is very obvious; but the model with
$\omega_0=-1.2<-1.0$, which is like the phantom field model, the
CMB spectrums have larger values. For the evolution of EoS is only
determined by the $\omega_0$, the value of it can be determined by
fitting the CMB observation. It is obvious that the recent
observations on the CMB power spectrums at large scale from WMAP
satellite have large error. The further results will depend on the
observation of the following WMAP and Planck satellites.

%From this figure, one can find that the smaller value of $\omega$
%follows smaller power spectrum, especially at large scale, which
%is for the dark energy with smaller $\omega$ can make the
%evolution of $\phi$ being faster. This feature is entirely same
%with the Universe with the scalar field dark energy.

Now let's return to the evolution of $\omega$. From Fig.[1], one
finds that $\omega$ crossing $-1$ can not be realized in these
models with a single YM field, no matter what values of $\omega_0$
we have chosen. For studying it more clear, assume the YM field
has an initial state of $|\omega+1|\ll1$, which follows that
$\beta\ll1$, the Eq.(\ref{16}) becomes
 \be
 \beta\propto a^{-2}.
 \ee
The value of $\beta$ will go to zero with the expansion of the
Universe. This means that $E$ will go to a critical state of
$E^2=\kappa^2$.  And the EoS is
 \be
 \omega+1\simeq\frac{4\beta}{3}\propto a^{-2}.\label{o+1}
 \ee
This result has two important characters: i) with the expansion of
the Universe, $\omega$ will go to the critical point of
$\omega=-1$. This is the most important character of this dark
energy model, which is very like the behavior of the vacuum energy
with $\omega\equiv-1$; ii) the value of $\omega>-1$ and
$\omega<-1$ all can realized, but it can not cross $-1$ from one
area to another. This character is same with the scalar field such
as the quintessence field, the k-essence and the phantom field
models.

It is interesting to ask: if these characters are correct just for
the YM model with the lagrangian as formula (\ref{L})? Whether or
not one can build a model, whose EoS can cross $-1$? So let's
consider the YM field model with a general effective lagrangian
as:
 \be
 \L_{eff}=G(F)F/2,
 \ee
where $G(F)$ is the running coupling constant, which is a general
function of $F$. If we choose $G(F)=b\ln|\frac{F}{e\kappa^2}|$,
this effective lagrangian returns to the from in Eq.(\ref{L}). The
dielectric constant also can be defined by
$\epsilon=2\partial\L_{eff}/\partial F$, which is
 \be
\epsilon=G+FG_{F}.
 \ee
Here $G_F$ represents $dG/dF$. We also discuss the homogeneous and
isotropy YM field with electric field $(B=0)$, then the energy
density and the pressure of the YM field are:
 \be
 \rho=E^2(\epsilon-\frac{G}{2}),
 \ee
 \be
 p=-E^2(\frac{\epsilon}{3}-\frac{G}{2})
 \ee
The energy density $\rho>0$ follows a constraint $G>-2FG_F$. The
EoS of this YM field is
 \be
 \omega=-\frac{3-2\gamma}{3-6\gamma},\label{omega}
 \ee
where we have defined that $\gamma\equiv\epsilon/G$. When the
condition of $\gamma=0$ can be got at some state with $E^2\neq0$
and $G(F)\neq0$, the state of $\omega=-1$ is naturally realized.
This condition can be easily satisfied. In the discussion as below
we only consider these kind of YM fields. For example, in the
model with the lagrangian (\ref{L}), $\gamma=0$ is got at the
state $E^2=\kappa^2$. Near this state, $\gamma>0$ leads to
$\omega<-1$, and $\gamma<0$ leads to $\omega>-1$. But if the YM
field has a trivial lagrangian with $G=constant$, which follows
that $\gamma\equiv1$, and $\omega\equiv1/3$. This is exactly the
EoS of the relativistic matter, and it can not generate the state
of $\omega<0$.
%It also
%means that only if the YM field has a running coupling constant,
%which is the function of $F$, the state of $\omega<-1/3$ can be
%realized. But this also is not correct for all condition, if
%$G(F)\propto F^\alpha$, where $\alpha$ is a constant number, then
%$\epsilon=(\alpha+1)G$, which makes
%$\omega=\frac{1-2\alpha}{6\alpha-3}$. This is also difficult to
%get a state of $\omega<-1/3$. In below discussion, we would not
%consider this condition.

To study the evolution of EoS, we also consider the YM equation,
which can be got by variation of $S$ with respect to
$A_{\mu}^{a}$,
 \be
 \partial_{\mu}(a^4\epsilon~
 F^{a\mu\nu})+f^{abc}A_{\mu}^{b}(a^4\epsilon~F^{c\mu\nu})=0,
 \label{F1}
 \ee
from the definition of $F^{a}_{\mu\nu}$, it is found that these
equations become a simple relation:
 \be
 \partial_{\tau}(a^2\epsilon E)=0,
 \ee
where $E$ is defined by $E^2=\Sigma_{i=1}^3E_i^2$. If
$\epsilon=0$, this equation is an identity, and from
(\ref{omega}), we know $\omega=-1$, which can't be differentiated
from cosmological constant. When $\epsilon\neq
 0$, this equation can be integrated to give
 \be
 a^2\epsilon E=constant.\label{k1}
 \ee
For we want to study whether or not the EoS of this YM field can
cross $\omega=-1$, here we assume its initial state is
$\omega\sim-1$. In this condition, from the expression of $p$ and
$\rho$, it follows that $\epsilon\sim0$,  $E$ and $G(F)$ nearly
keep constant, which is for the Universe is nearly de Sitter
expansion and $\rho\sim-G(F)E^2/2$ is nearly a constant in this
Universe. So the YM equation suggests that
 \be
 \epsilon\propto a^{-2}.
 \ee
From the EoS of (\ref{omega}), one knows that
 \be
 \omega+1\propto a^{-2}.\label{o1+1}
 \ee
This is the EoS evolution equation of the general YM field dark
energy models. It is exactly same with special case of
Eq.(\ref{o+1}). So it also keeps the characters of the special
case with the lagrangian (\ref{L}): $\omega$ will run to the
critical point $\omega=-1$ with the expansion of the Universe. But
it can not cross this critical point. These is the general
characters of these kind of YM field dark energy models. For
showing this more clear, we discuss two example models.

First we consider the YM field with the running coupling constant
 \be
 G(F)=B(F^n-F_c^n),
 \ee
where $B$ and $F_c$ are quantity with positive value, and $n$ is a
positive number. The constraint of $\rho>0$ follows that
 \be
 F>\frac{F_c}{\sqrt[n]{1+2n}}.
 \ee
The dielectric constant can be easily get
 \be
 \epsilon=G+FG_F=B(n+1)F^n-BF_c^n,
 \ee
and
 \be
 \gamma=(n+1)+\frac{nF_c^n}{F^n-F_c^n}.
 \ee
It is obvious that when $F=F_c/\sqrt[n]{n+1}$, $\gamma=0$ is
satisfied, and which leads to $\omega=-1$. Near this critical
state, $E\sim\sqrt[2n]{F_c^n/(n+1)}$. So the YM equation of
(\ref{k1}) becomes
 \be
\frac{An}{n+1}\sqrt[2n]{\frac{F_c^{3n}}{n+1}} \gamma\propto
 a^{-2},
 \ee
which follows that $\gamma\propto a^{-2}$. From the expression of
$\omega$ in Eq.(\ref{omega}), one can easily get
 \be
 \omega+1\simeq-\frac{4\gamma}{3}\propto a^{-2}.
 \ee
This is exact same with the evolution behavior shown in formula
(\ref{o1+1}).

Another example, we consider the YM field with the coupling
constant of
 \be
 G(F)=1-\exp(1-\frac{F}{F_c}),
 \ee
where the constant quantity $F_c\neq0$. When $F\gg F_c$, this
lagrangian becomes the trivial case with $G(F)=1$, but when $F$ is
near $F_c$, the nonlinear effect is obvious. Then
 \[
 \epsilon=1+(\frac{F}{F_c}-1)\exp(1-\frac{F}{F_c}).
 \]
so the critical state of $F=0.433F_c$ leads to $\gamma=0$ and
$\omega=-1$. By the similar discussion, from the YM field
(\ref{k1}), one can also get $\gamma\propto a^{-2}$ near this
critical state, which generates $\omega+1\propto a^{-2}$.

%%%%%%%%%%%%%%%%%%%%%%%%%%%

\section{ Two YM Fields Model}

In the former section, we have discussed that the dark energy
models with single YM field can't form a state of $\omega$
crossing $-1$, not matter what kind of lagrangian or initial
condition. But we should notice another character: the YM field
has the EoS of $\omega\propto-1+ a^{-2}$, when its initial value
is near the critical state of $\omega=-1$. So if the YM field has
an initial state of $\omega>-1$, it will keep this state with the
evolution of the Universe, which is like the quintessence models.
But if its initial state is $\omega<-1$, it will also keep it,
which is like the phantom models. This makes that we can build a
model with two different free YM fields, one having an initial
state of $\omega>-1$ and the other being $\omega<-1$. In this kind
of models, the behavior of $\omega$ crossing $-1$ is easily got.
This idea is like the quintom models\cite{quintom}, where the
authors built the model with a quintessence field and a phantom
field.

In the below discussion, we will build a toy example of this kind
of model. Assume the dark energy is made of two YM fields with the
effective lagrangian as Eq.(\ref{L})
 \be
 \L_{i}=\frac{b}{2}F_i\ln|\frac{F_i}{e\kappa_i^2}|,~~~(i=1,2)
 \ee
where $F_i=E_i^2 (i=1,2)$, and $\kappa_1\neq\kappa_2$. Their
dielectric constants are
 \be
 \epsilon_i\equiv\frac{2\partial\L_{i}}{\partial F_i}=b\ln|\frac{F_i}{\kappa_i^2}|.
 \ee
From the YM field kinetic equations, we also can get the
relations:
 \be
 a^2\epsilon_iE_i=C_i,\label{kin}
 \ee
where $C_i (i=1,2)$ are the integral constant, which are
determined by the initial state of the YM fields. If the YM field
is a phantom like field with $\omega_i<-1$, then $\epsilon_i<0$
and $C_i<0$. At the same time, a quintessence like YM field
follows that $C_i>0$. Here we choose the YM field of $\L_1$ as the
phantom like field with $C_1<0$, and $\L_2$ as the quintessence
like field with $C_2>0$. The energy density and pressure are
 \be
 \rho_i=\frac{E_i^2}{2}(\epsilon_i+b),~~p_i=\frac{E_i^2}{2}(\frac{\epsilon_i}{3}-b),
 \ee
so the total EoS is:
 \be
 \omega\equiv\frac{p_1+p_2}{\rho_1+\rho_2}=\frac{E_1^2(\frac{\beta_1}{3}-1)+E_2^2(\frac{\beta_2}{3}-1)}
 {E_1^2(\beta_1+1)+E_2^2(\beta_2+1)},
 \ee
where we have also defined that $\beta_i\equiv\epsilon_i/b$. Using
the relation of $\beta_i$ and $E_i$, we can simplify the equation
of state as
 \be
 \omega+1=\frac{4}{3}\frac{e^{\beta_1}\beta_1\alpha+e^{\beta_2}\beta_2}
 {e^{\beta_1}(\beta_1+1)\alpha+e^{\beta_2}(\beta_2+1)},\label{ome}
 \ee
where $\alpha\equiv\kappa_1^2/\kappa_2^2$. We need this dark
energy has the initial state of $\omega>-1$, which requires that
the field of $\rho_2$ is dominant at the initial time. This is
easily obtained as long as at this time
$E_1^2(\beta_1+1)<E_2^2(\beta_2+1)$ is satisfied. The finial
state, we need to get $\omega<-1$, which means that $\rho_1$ is
dominant, and the behavior of crossing $-1$ realized in the
intermediate time. But how to get this? From the before
discussion, we know in the Universe with only one kind of YM field
(i=1 or 2), the YM equation follows that $\epsilon_i\propto
a^{-2}$. And it will go to the critical state $\epsilon_i=0$ with
the expansion of the Universe. At this state, $E_i^2=\kappa_i^2$,
and $\rho_i=bE_i^2/2=b\kappa_i^2/2$ keeps constant. So in this two
YM fields model, if we choose the condition
$\kappa_1^2>\kappa_2^2~ (\alpha>1)$, this may follow that the
finial energy density $\rho_1>\rho_2$, and $\rho_1$ is dominant.
%If it has a
%initial state of $\omega_1<-1$, which will make the total EoS
%$\omega<-1$. The initial condition of $\rho_1<\rho_2$ is also
%easily got, which only needs $E_1^2(\beta_1+1)<E_2^2(\beta_2+1)$.
For this intent, we build this model with the condition as below:
choosing $\alpha=1.5$, which can ensure the finial state, the
first kind of YM field $(i=1)$ is the dominant matter. At the
present time, corresponding to the scale factor $a_0=1$, we choose
that $\beta_1=-0.4<0$ and $\beta_2=0.2>0$, which keeps that the
first field always having a state of $\omega_1<-1$ (like the
phantom) and the second field with $\omega_2>-1$ (like the
quintessence). These choice of $\beta_i$ leads to the present EoS
 \be
 \omega=-1+\frac{4}{3}\frac{e^{\beta_1}\beta_1\alpha+e^{\beta_2}\beta_2}
 {e^{\beta_1}(\beta_1+1)\alpha+e^{\beta_2}(\beta_2+1)}=-1.10<-1,
 \ee
which is like the phantom field. Since $\rho_1$ increases, and
$\rho_2$ decreases with the expansion of the Universe, there must
exist of a time, before which $\rho_2$ is dominant, and this leads
to the total EoS $\omega>-1$ at that time.

Combining the Eqs.(\ref{kin}) and (\ref{ome}), we can solve the
evolution of EoS $\omega$ with the scale time in numerical
calculation, where the relation of $C_1$ and $C_2$ is easily got
 \[
 \frac{C_1}{C_2}=\frac{\beta_1e^{\beta_1/2}}{\beta_2e^{\beta_2/2}}=-1.48.
 \]
For each kind of YM field, its EoS is
 \be
 \omega_i=\frac{\beta_i-3}{3\beta_i+3}~~~(i=1,2).
 \ee
The condition of $\beta_i>0~(\beta_i<0)$ will generate
$\omega_i>-1~(\omega_i<-1)$. The evolution of them are shown in
Fig.[3]. This is exact result we expect, $\beta_i$ runs to the
critical point of $\beta_i=0$ with the expansion of the Universe,
which makes $\omega_i$ runs to $\omega_i=-1$, no matter what kind
of initial values. This is same with the single YM field model.
With $\beta_i\rightarrow 0$, the strength of the field $E_i^2$
will also go to its critical point $E_i^2=\kappa_i^2$. This can be
shown in Fig.[4]. For we have chosen the condition of
$\alpha\equiv\kappa_1^2/\kappa_2^2>1$, it must lead to
$E_1^2>E_2^2$ at some time, the first kind of YM field becomes
dominant, and the total EoS $\omega<-1$ is realized. This can be
shown in Fig.[1]. In Fig.[2], we also plot the CMB power spectrum
in the Universe with this kind of YM field dark energy, and find
that it is difficult to be distinguished from the $\Lambda$CDM
Universe, which is for the effect of the dark energy on the CMB
power spectrum is an integral effect from CMB decoupling time to
now. But the evolution detail of $\omega$ is not obvious. This is
the disadvantage of this way to detect dark energy.

Now, let's conclude this dark energy model, which is made of two
YM fields. One has the EoS of $\omega_1<-1$ and the other is
$\omega_2>-1$. At the initial time, we choose their condition to
make $\rho_1<\rho_2$, and second kind of YM field is dominant,
which makes the total EoS $\omega>-1$ at this time. This is like
the quintessence model. For the $\omega_1<-1$ keeps for all time,
from the Friedmann equations, one knows that its energy density
will enhance with the expansion of the Universe. And at last it
will run to its critical point $\rho_1=b\kappa_1^2/2$. And the
same time, $\rho_2$ will decrease to its critical state
$\rho_2=b\kappa_2^2/2$. For we have chosen
$\kappa_1^2/\kappa_2^2>1$, which must make $\rho_1=\rho_2$ and
some time, and after this, $\rho_1$ is dominant, the total EoS
$\omega<-1$. So the equation of state $\omega$ crossing $-1$ is
realized. It is simply found that this kind of crossing must be
from $\omega>-1$ to $<-1$, which is exactly same with the
observations. But the contrary condition, from $\omega<-1$ to
$>-1$ can't be realized in this kind of models.

%%%%%%%%%%%%%%%%%%%%%%%%%%

\section{Conclusion and Discussion}

In summary, in this letter we have studied the possibility of
$\omega$ crossing $-1$ in the YM field dark energy models, and
found that the single YM field models can not realize, no matter
what kind of their effective lagrangian, although this kind of
models can naturally give a state of  $\omega>-1$ or $\omega<-1$,
which depends on their initial state. Near the critical state of
$\omega=-1$, the evolution of their EoS with the expansion of the
Universe is same, $\omega+1\propto a^{-2}$, which means that the
Universe will be a nearly de Sitter expansion. This is the most
attractive character of this kind of models, and this makes it
very like the cosmological constant. So the Big Rip is naturally
avoided in this model. But this evolution behavior also shows that
the single field models can not realize $\omega$ crossing $-1$.
This is same with the single scalar field models.

But in these models, $\omega>-1$ and $\omega<-1$ all can be easily
got. The former behavior is like a quintessence field, and the
later is like a phantom field. So one can build a model with two
YM fields, and one field with $\omega<-1$ and the other with
$\omega>-1$. This idea is very like the quintom models. Then we
give an example model and find that in this model, the property of
crossing the cosmological constant boundary can be naturally
realized, and we also found that this crossing must be from
$\omega>-1$ to $<-1$, which is exact the observation result. In
this model, the state will also go to the critical state of
$\omega=-1$ with the expansion of the Universe, as the single YM
field models. This is the main character of the YM field dark
energy models, which makes the Big Rip is avoided. The present
models we discuss in this paper are in the almost standard
framework of physics, \emph{e.g}. in general relativity in
4-dimension. There does not exist phantom or higher derivative
term in the model, which will lead to theory problems in field
theory. Instead, the YM field as (\ref{L}), is introduced, which
includes the gauge invariance, the Lorentz invariance, the correct
trace anormaly, and the asymptotic freedom. These are the
advantages of this kind of dark energy models. But these models
also exist some disadvantages: first, what is the origin of the YM
field? and why its renormalization scale $\kappa^2$ is so low as
the present density of the dark energy? In the two YM fields
model, we must choose $\alpha>1$ to realize the $\omega$ crossing
$-1$, which is a mild fine-tuning problem. All these make this
kind of models being unnatural. These are the universal problems
which exist in most dark energy models.  If considering the
possible interaction between the YM field and other matter,
especially the dark matter, which may have some new
character\cite{zhang2}. This topic had been deeply discussed in
the scalar field dark energy models\cite{inter}, but had not been
considered in this paper.

%%%%%%%%%%%%%%%%%%%%%%%%%%%%%%%%%%%%%%%%%%%%%%%%%%%%%%%%%%%%%%%%%%%%%%%%%%%%%%%%%%%
%%%%%%%%%%%%%%%%%%%%%%%%%%%%%%%%%%  Acknowledgments   %%%%%%%%%%%%%%%%%%%%%%%%%%%%%%%%%%%%%%%
%%%%%%%%%%%%%%%%%%%%%%%%%%%%%%%%%%%%%%%%%%%%%%%%%%%%%%%%%%%%%%%%%%%%%%%%%%%%%%%%%%%

\section*{Acknowledgements}
Y. Zhang's research work has been
supported by the Chinese NSF (10173008) and by NKBRSF (G19990754).

%%%%%%%%%%%%%%%%%%%%%%%%%%%%%%%%%%%%%%%%%%%%%%%%%%%%%%%%%%%%%%%%%%%%%%%%%%%%%%%%%%%
%%%%%%%%%%%%%%%%%%%%%%%%%%%%%%%%%%  BIBLIOGRAPHY  %%%%%%%%%%%%%%%%%%%%%%%%%%%%%%%%%%%%%%%%
%%%%%%%%%%%%%%%%%%%%%%%%%%%%%%%%%%%%%%%%%%%%%%%%%%%%%%%%%%%%%%%%%%%%%%%%%%%%%%%%%%%

\newpage

 \begin{figure}
 \centerline{\includegraphics[width=11cm]{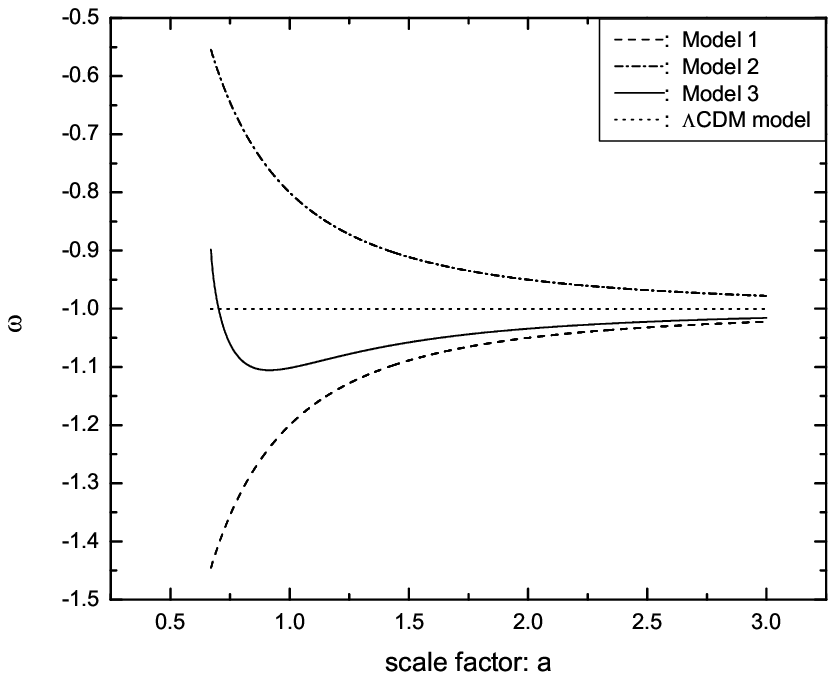}}
 \caption{\small The evolution of the state equation $\omega$ with the scale factor
 $a$. The model 1 denotes the single YM field model with the
 present EoS $\omega_0=-1.2$, model 2 denotes the model with
 $\omega_0=-0.8$, and model 3 denotes the two YM fields model, where we
 have set $\kappa_1^2=1.5\kappa_2^2$, and $\beta_1=-0.4$,
 $\beta_2=0.2$ at present with $a_0=1$.}
 \end{figure}

 \begin{figure}
 \centerline{\includegraphics[width=11cm]{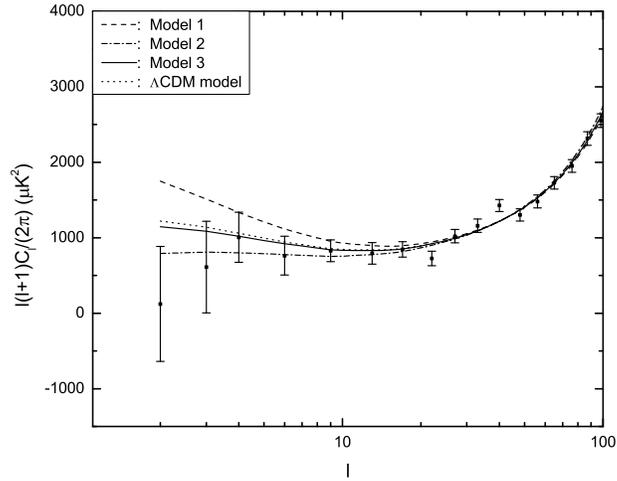}}
 \caption{\small The CMB anisotropy power spectrum.
 The model 1 denotes the single YM field model with the
 present EoS $\omega_0=-1.2$, model 2 denotes the model with
 $\omega_0=-0.8$, and model 3 denotes the two YM fields model, where we
 have set $\kappa_1^2=1.5\kappa_2^2$, and $\beta_1=-0.4$,
 $\beta_2=0.2$ at present with $a_0=1$. In this figure, the dots denote
 observation results from the WMAP\cite{map} satellite.
 Here we have used the CMBFAST program\cite{cmbfast}.}
 \end{figure}

 \begin{figure}
 \centerline{\includegraphics[width=11cm]{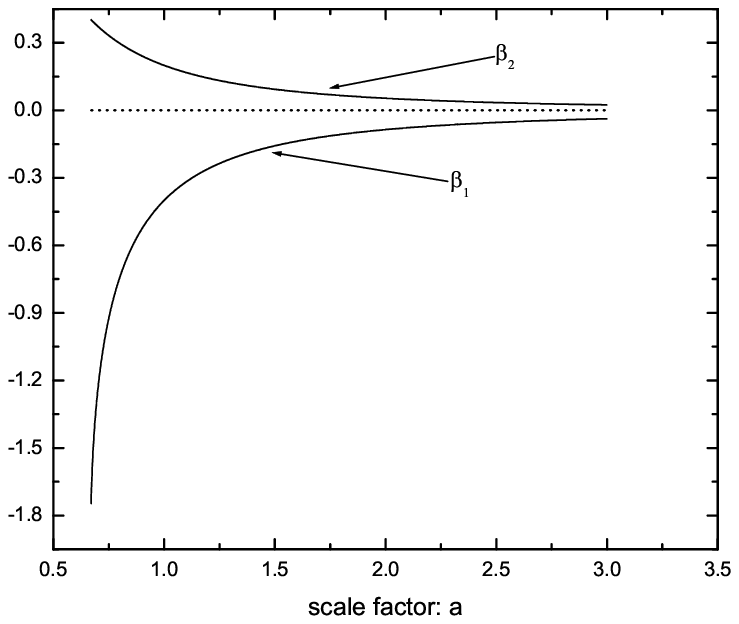}}
 \caption{\small The evolution of the $\beta_1$ and $\beta_2$ with the scale factor $a$,
 where we have set $\kappa_1^2=1.5\kappa_2^2$, and $\beta_1=-0.4$,
 $\beta_2=0.2$ at present time with $a_0=1$.}
 \end{figure}

 \begin{figure}
 \centerline{\includegraphics[width=11cm]{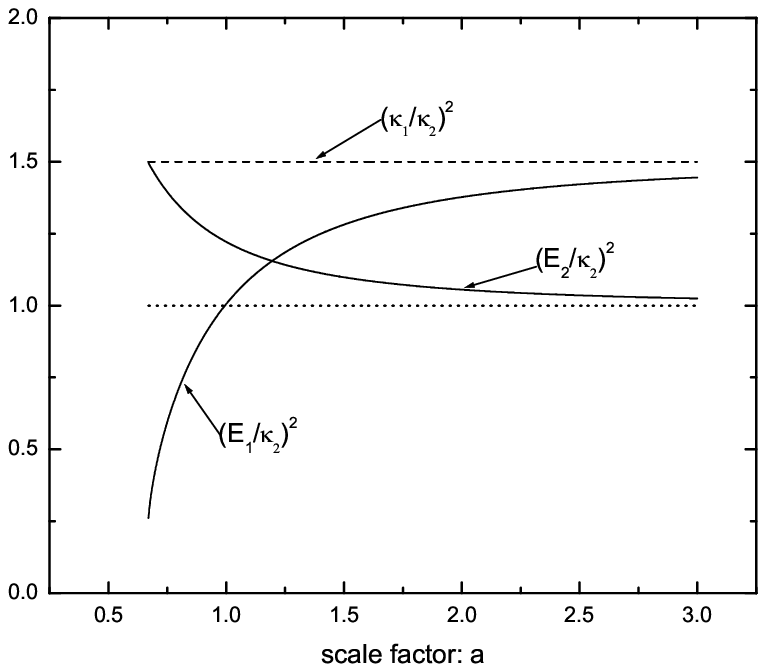}}
 \caption{\small The evolution of the $E_1$ and $E_2$ with the scale factor $a$,
 where we have set $\kappa_1^2=1.5\kappa_2^2$, and $\beta_1=-0.4$,
 $\beta_2=0.2$ at present with $a_0=1$.}
 \end{figure}

\end{document}